\documentclass[]{aa}

\usepackage{graphicx}
\usepackage{txfonts}
\usepackage{natbib}
\bibpunct{(}{)}{;}{a}{}{,}

\def\pcite#1{[ref]}

\begin{document}
   \title{Quantifying Chaos in Models of the Solar Neighbourhood}
   \subtitle{}
   \author{Dalia Chakrabarty \inst{1}
           \and Ioannis V. Sideris  \inst{2}
          }
   \offprints{Dalia Chakrabarty}
   \institute{
              School of Physics $\&$ Astronomy,
              University of Nottingham, 
              Nottingham NG7 2RD, U.K.
              \email{dalia.chakrabarty$@$nottingham.ac.uk}
              \and 
	      Institute for Theoretical Physics, University of Z\"urich,
 		  Winterthurerstrasse 190, 
		  Z\"urich CH-8057, Switzerland.
              \email{sideris@physik.unizh.ch}
             }

   \date{\today}

\abstract {} {To quantify the amount of chaos that exists in the local
phase space.} {A sample of orbits from four different models of the
Solar neighbourhood phase space are analysed by a new chaos
identification (and quantification) technique. While three of the used
models bear the signature of the perturbation due to both the Galactic
bar and the spiral pattern, the last of the models is a bar only
one. We explore the models by inter-comparing the corresponding values
of chaos strength that is induced at the various energy levels .}{(1) We
find that of all the viable models that have been demonstrated to
successfully reproduce the local phase space structure, i.e. those
that include the bar as well as the spiral, bear strong chaoticity,
though the model that implies the highest degree of chaos is the one
in which overlap of the major resonances of the bar and the spiral
occurs. The bar only model is found to display regularity. (2) We advance
chaos to be primarily responsible for the splitting of the
Hyades-Pleiades mode (the larger mode) of the local velocity
distribution}{}

 \keywords{chaos --
                Galaxy --
                solar neighbourhood
               }

\maketitle

\section{Introduction}
\noindent
The availability of transverse velocities of nearby stars from {\it
Hipparcos}, facilitated the construction of the local phase space
distribution \citep{fux, skulljan, walterdf}. In contradiction to the
conventional idea of stellar dynamics, all representations of this
distribution manifest strong non-linearity and multi-modalness. This
clumpy nature of the solar neighbourhood velocity distribution ($f$)
has been addressed in \citep{fux, quillen, walterolr, chakrabarty_aa,
famaey, tremainewu} and others; consensus appears to be emerging as to
the origin of the basic bimodal nature of the distribution in terms of
scattering off the Outer Lindblad Resonance of the Galactic bar
($OLR_b$).

However, a dynamical basis for the existence of the other structure
(such as the Hyades, Pleiades, Sirius, Coma Berenicus stellar streams)
has attracted less of a focus. \cite{chakrabarty_aa} (hereafter,
Paper~I) concluded the observed phase space structure to be due to the
dynamical influence of the Galactic bar and 4-armed spiral pattern;
the influence of the bar alone, or the spiral alone were reported to
be insufficient in explaining the present day observations of the
solar neighbourhood. \citet{quillen} invoked the chaos caused by the
overlapping of the $OLR_b$ and the 4:1 resonance of the Galactic
spiral pattern to explain the chaos dominated state of the local disk,
a ramification of which, it was suggested, is the clumpy nature of
$f$.

In spite of these investigations, the quantification of the degree of
chaos in the solar vicinity, has not been undertaken yet. This is of
interest in interpreting the state of the local patch in the disk and
extrapolate this notion to the understanding of the Galactic disk as a
whole as well as of outer disks in external spiral systems. The former
of these motivations is to get a boost in the near future, with the
quantity and spatial cover promised in the data from the upcoming {\it
GAIA} mission. Here we present a new technique for estimating the
amount of chaos that is induced in the solar neighbourhood, by the
Galactic bar and spiral pattern. The different models used in Paper~I
will be analysed by the technique advanced in \citet{sideris1}. Thus,
the aim of this paper is to evaluate the extent of chaos in the solar
neighbourhood and examine the possible connection between the
identified chaos and the local phase space structure.

This paper is organised as follows. The following section deals with
the models, while in \S3 the equations of motion are briefly
discussed. The chaos quantification technique is advanced in \S4 and
the recovered results are presented herein. \S5 is dedicated to a
discussion of some aspects of the work. The paper is rounded off with
the concluding remarks in \S6.

\begin{figure}
\includegraphics[width=7cm]{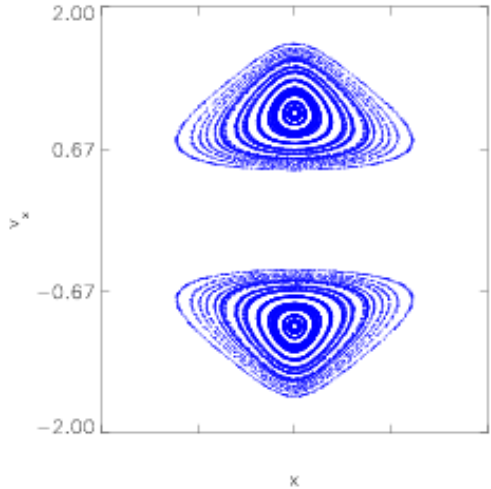}
\caption{\label{fig:fig1.ps} Poincare section for the bar-only model,
at the energy of -0.75. The white inner regions mark the part of
$x-v_x$ space that is not populated by orbits for the specific
implementation of our numerical experiment.  The blue lines are
invariant curves, (i.e curves representing the 4-d regular orbits in
the 2-dimensional Poincare space).}
\end{figure}

\begin{figure}
\includegraphics[width=7cm]{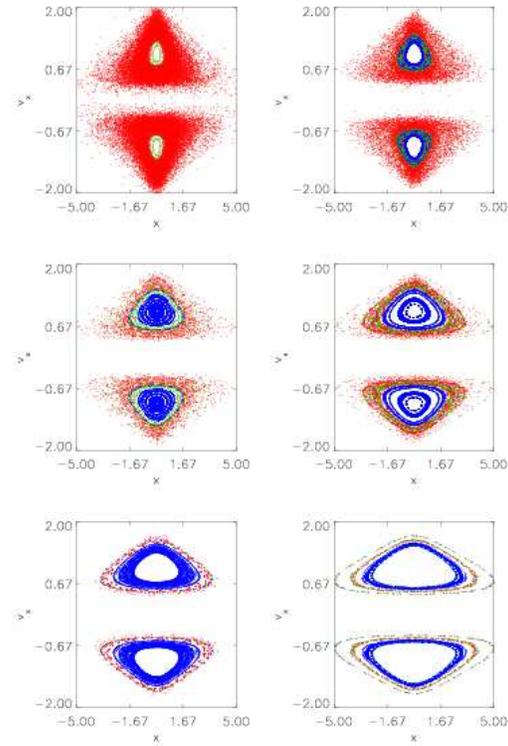}
\caption{\label{fig:fig2.ps} Surface of sections of orbits integrated
in the model with spiral to bar pattern speed ratio of 21/55. Red
signifies strong chaos, green signifies weak chaos and blue signifies
regularity. Each panel represents a surface of section plot for a
particular energy value; top left panel corresponds to $J$=-0.300, top
right to -0.5, middle left to -0.75, middle right to -1.0, bottom left
to -1.25 and bottom left to orbits corresponding to energy of -1.5. It
is evident that chaos decreases as energy decreases.}
\end{figure}

\section{Models of the Local Disk}
\noindent
As said before, here we endeavour to infer the degree of chaos present
in the vicinity of the Sun by gauging chaoticity of solar
neighbourhood models that were presented in Paper~I. Thus, the
justification of the choice of the relevant parameters will not be
repeated here; rather, it is the aspect of quantification of the chaos
inherent in each of these models that we discuss below.

In Paper~I, an annulus in the outer part of the Galactic disk was
modelled by test particle simulations, in which a warm exponential
disk was stirred by the bar or a spiral pattern alone, or by both
these perturbations jointly. In these simulations, the Galactic disk
is assumed to be ideal with the disk stars assumed to be drawn from a
4-D phase space. A sample of phase space coordinates were chosen from
a model initial phase space distribution function (chosen to ensure an
exponential surface mass density profile and enough warmth to attain
the velocity dispersions and vertex deviation observed in the solar
neighbourhood today). These coordinates were allowed to evolve with time in the
presence of the potential of the disk and the perturbation(s),
i.e. the bar or (and) spiral. The bar was modelled as a rigidly
rotating quadrupole (see Equation~1 in Paper~I) with a perturbation
strength that is half the strength of the bar used in Fux, 2001. The
spiral pattern is modelled as a logarithmic spiral that is 4-armed
\citep{vallee_02} and tightly wound (pitch angle of 15$^\circ$), as
the model spiral pattern used by \cite{johnston01}; this choice of
number of arms and a low pitch angle also ties in with the suggestion
of \cite{melnik, bissantz, peter, vallee_02}. The initial disk
configuration is characterised by a logarithmic potential to ensure
flat rotation curve and a doubly-cut out distribution function
\citep{wynjenny} that ensures an exponential surface stellar mass
density profile. This distribution function is parametrised by a
hotness parameter that is maintained sufficiently high to ensure the
recovery of velocity dispersions and vertex deviation that match with
the observed values of these quantities in the solar neighbourhood
today.

The orbits were recorded in the annulus between $R=1.7R_{\rm CR}$ to
$R=2.3R_{\rm CR}$, where $R_{\rm CR}$ is the corotation radius of the
bar; $OLR_b$ occurs at 1.7$R_{\rm CR}$ for the above mentioned choice
of the disk potential. In this work, all lengths are expressed in
units of $R_{\rm CR}$ and given the scale free nature of our disk
configuration, the physical value of $R_{\rm CR}$ is not relevant. An
important parameter that was varied to define the individual models is
the ratio between the pattern speeds of the bar ($\Omega_b$) and the
4-armed spiral ($\Omega_s$). In every other respect, the bar+spiral
models are similar to each other. The bar-only model is similar to the
bar+spiral models in every respect except that there is no
perturbation from the spiral in this model. Likewise for the
spiral-only model. Thus, the 5 models used in Paper~I are:
\begin{itemize}
\item bar alone perturbing disk.
\item bar and spiral acting in concert with $\Omega_b/\Omega_s$=55/25.
\item bar and spiral acting in concert with $\Omega_b/\Omega_s$=55/21.
\item bar and spiral acting in concert with $\Omega_b/\Omega_s$=55/18.
\item spiral acting alone.
\end{itemize}
From Paper~I we learn that out of these 5 models, the first four were
found to give rise to phase space structure that is reminiscent of the
observed structure (checked via a hypothesis testing technique),
though the bar-only model was rejected on further dynamical
grounds. In particular, the bar+spiral model that is characterised by
$\Omega_b/\Omega_s$=55/21 is the one that ensures that the ILR of the
spiral ($ILR_s$) occurs at the same physical location as the
$OLR_b$. Thus, this is the model that corresponds to overlap of the
major resonances of the two perturbations therefore augers interesting
dynamical consequences.

\section{Equations of Motion}
\noindent
In this section, the stellar equations of motion are discussed. Below
is presented the Hamiltonian in an inertial frame, in galactocentric
coordinates $x_i$, for i=1,2 and their conjugate momentum (or velocity
$v_i$), given the logarithmic potential of the background disk
($\sim\ln(R)$, where $R=\sqrt{x_1^2+x_2^2}$) and the perturbations
due to the quadrupolar bar ($\Phi_{bar}$) and the logarithmic $m$=4
spiral pattern ($\Phi_{spiral}$).
\begin{equation}
{\cal {H}} = \displaystyle{\sum_1^2 v_i^2 +\ln(R) + \Phi_{bar} + \Phi_{spiral}}, 
\end{equation}
where the potential of the bar and the spiral in our scale-free units
(i.e, all lengths are expressed in units of the bar corotation
radius), in the inertial frame, at time $t$ are:
\begin{eqnarray}
\Phi_{bar} &=& -\epsilon_{bar}\displaystyle{\frac{\cos2(\phi-{\Omega_b}t)}{R^3}} \nonumber \\ 
\Phi_{spiral} &=& -\epsilon_{spiral}K(\alpha,m)\displaystyle{e^{i[m(\phi-{\Omega_s}t)]}R^{i\alpha-\frac{1}{2}}}.
\label{eqn:pots}
\end{eqnarray}
Here $\alpha=m\:{\cot}(i)$, where $i$ is the pitch angle of the spiral
and $m$ is the number of arms in the pattern ($i$=15$^\circ$ and $m$=4
for our models). $K(\alpha, m)$ is the Kalnajs gravity function as
defined in Equation~13 of \cite{chakrabarty_04}. Also,
$\epsilon_{bar}$ and $\epsilon_{spiral}$ are the bar and spiral
strengths, defined in terms of the fractional contribution of the
particular perturbation to the field due to the background disk
($\approx$3.6$\%$ for the bar and the spiral). Lastly, here $\phi$ is
the azimuthal coordinate: $\phi=\tan^{-1}(x_2/x_1)$. See Section~2.2 of
Chakrabarty (2004) for a detailed discussion of the equations of
motion.

Thus, in the inertial frame, the equations of motion are:
\begin{equation}
\ddot{\bf{x_i}} = \displaystyle{\frac{-\bf{x_i}}{R^2} - \nabla\Phi_{bar} - \nabla\Phi_{spiral}}.
\end{equation}
When the only imposed perturbation is the bar, recording the orbits in
the frame of the bar implies that the Jacobi Integral is:
\begin{equation}
{\cal{H_J}} = \displaystyle{\sum_1^2 v_{i}^2 +\ln(R) -\epsilon_{bar}\frac{\cos2\phi}{R^3}}.
\end{equation}
Thus, in this case, ${\cal{H_J}}$ is an integral of motion and the
surfaces of section that are recovered for this 4-D phase space, by
setting $v_y$=0, is two dimensional. 

However, in the multiple pattern speed scenario, the Hamiltonian is no
longer the Jacobi Integral; thus, when the spiral pattern is included
as the second perturbation, and the orbits recorded in the frame
rotating with the bar, the orbital energy is:
\begin{eqnarray}
{\cal{H_J}} =& \displaystyle{\sum_1^2 v_i^2 +\ln(R) -\epsilon_{bar}\frac{\cos2(\phi)}{R^3}} \nonumber \\ 
 &\displaystyle{-\epsilon_{spiral}K(\alpha,m)e^{i[m(\phi-{(\Omega_s-\Omega_b)}t)]}R^{i\alpha-\frac{1}{2}}}.
\end{eqnarray}
It is obvious
that the quantity ${\cal{H_J}}$ in Equation~5 returns to the same value
periodically for period $T=m*\pi/(\Omega_b-\Omega_s)$, so if data are
recorded stroboscopically every such period, ${\cal{H_J}}$ is equivalent
to an integral of motion. Then out of the recorded points
per orbit (which are recorded only when $t=T$) one can construct
two-dimensional surfaces of section by employing a second
constraint, in our case by choosing to plot only the points
which have $v_y$=0.  Any other constraint one may impose, e.g.
$v_x$=0, should give the same results regarding the
percentages of chaotic and regular orbits or strengths of
chaos, since we quantify the same set of orbits but at a
different surface section.

\section{Quantification of Chaos and Results}
\noindent
The quantification of chaos for the orbits of this paper was achieved
by the use of a new measure first introduced by
~\citet{sideris1}. This technique is based on the recognition of
smooth patterns in the signals associated with an orbit. It was shown
in the original paper that the extrema of regular orbits correlate in
such ways so to build smooth curves. This inherent smoothness,
typically hidden inside the signal, can then be implemented to define
a measure of regularity, through an intricate interpolation
scheme. The simple picture is that the smoother the curves the more
regular the signal is.

A chaotic orbit usually evolves in a divided phase space (a phase
space which is characterised by both regular and chaotic
regions~\cite{contopoulos1, sideris2}). In such a regime, any
chaotic orbit (provided it is integrated for long enough timescales)
will experience two kinds of dynamical epochs: strong or wild chaos
and weak or sticky chaos~\cite{shirts1, contopoulos1}. Strong
chaos is associated with motion of the orbit far away from the regular
islands. Such motion is completely unpredictable, and the chaotic
orbit attempts to cover broad parts of the chaotic sea energetically
available to it. When the orbit moves close to the regular islands it
becomes trapped for a long time around them, in practice,
attempts to mimic regularity. The closer to a regular island the
chaotic orbit moves the more persuasive this mimicry is.

\begin{figure*}
\includegraphics[width=15cm]{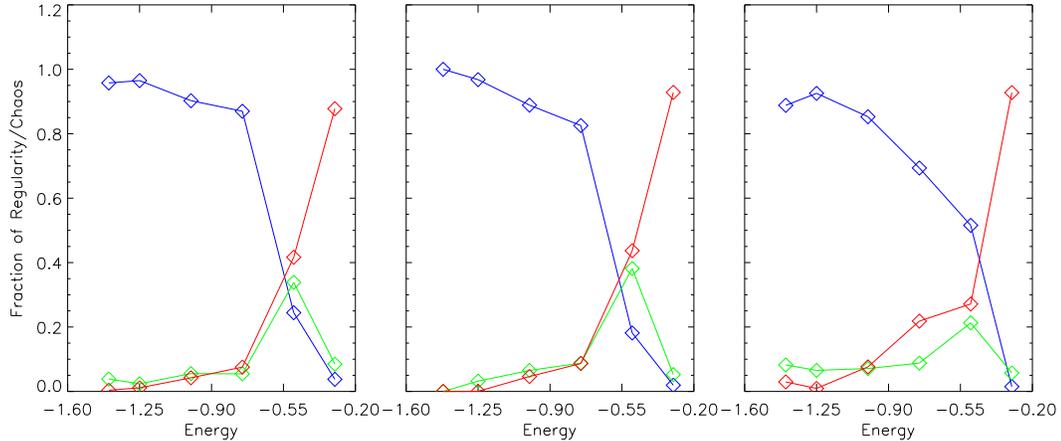}
\caption{\label{fig:fig3.ps} Fractions of chaotic orbits (in red),
weakly chaotic orbits (in green) and regular orbits (in blue), plotted
as functions of energy, for the three bar+spiral models 18/55 (left),
21/55 (middle), 25/55 (right).  }
\end{figure*}

The patterns method can identify when an orbit gets into weakly
chaotic regimes. Semi-smooth curves correlating extrema of the signal
of the orbit appear in that epoch of its evolution. The big advantage
of the pattern method is that it treats orbits as sets of segments,
piece by piece, and not as one monolithic entity as other measures
typically do. This is how it achieves to distinguish parts of the
orbit where weak chaos is experienced.

We applied this method to the orbits associated with the
aforementioned simulations. For every simulation a number of orbits
that correspond to a given value of energy, were randomly extracted in
several different energy bands and the chaos quantification followed.

For the bar only model for the six energies evolved (from -0.3 to
-1.5) no chaos was found. In Figure~\ref{fig:fig1.ps}, we show a
surface of section that is constructed for orbits characterised by an
energy of -0.75. All the surfaces of sections presented herein are
recorded for the orbits crossing the plane $v_y=0$.

The results for six different energies for the ratio 21/55 can be seen
in Figure~\ref{fig:fig2.ps}. Similar pictures hold true for models
18/55 and 25/55. In all three models it is obvious that chaos is very
strong for large energies but reduces as energy decreases.

To compare the chaos inducing ability of the different models, the
fraction of the regular and (strongly and weakly) chaotic orbits is shown in
Figure~\ref{fig:fig3.ps}. These plots show the percentage of
chaotic orbits appearing in the three models. One may notice that the
case 21/55 is quantified as more chaotic than the other cases.

\begin{figure}
\includegraphics[width=7cm]{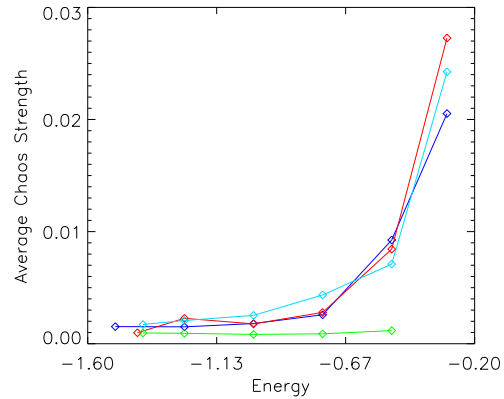}
\caption{\label{fig:fig4.ps} 
Average strength of chaos against energy, for the four models 18/55,
21/55, 25/55 and bar only. Blue signifies the 18/55 model, red 21/55,
cyan 25/55 and green the bar only model. }
\end{figure}

In Figure~\ref{fig:fig4.ps} the chaos strength is plotted with respect to
the energy for the four models $\Omega_s/\Omega_b$=18/55, 21/55, 25/55
and the bar-only.

\section{Discussions}
\noindent
Our chaos quantification technique helps shed light on the models; we
find that at the higher energies, {\it the model that manifests the
highest chaos is the model that ensures resonance overlap} (the 21/55
model). This is in line with our expectations of course, but it is
also interesting to note that the {\it chaos induced by the other
bar+spiral models is not much less either}. At the same time, from
Paper~I, we know that all three of the bar+spiral models
were successful in explaining the observed structure of the local
phase space. This adds weight to the suggestion that {\it chaos is
responsible for the clumps of the local velocity space}. (Of course,
this is only part of the story, since scattering off the Outer
Lindblad Resonance of the bar and the effects of minor resonances of
the bar and the spiral are also important, as reported in Paper~I).

To understand the trends in our results, we need to invoke the
following: the ILR of the spiral pattern is an "angular momentum
emitter" \citep{kalnajs_bell}, the basic effect of which is to "stir
without heating" \citep{binney_sellwood}. This idea that the Inner
Lindblad Resonance of the spiral ($ILR_s$) is the location from which
stars are driven outwards, is corroborated by the experiments of
\cite{chakrabarty_04}. Now in our modelling, we choose to record our
orbits in an annulus that extends from $R=$1.7$R_{CR}$ to 2.3$R_{CR}$,
where $R_{CR}$ is the corotation radius of the bar. So the occurrence
of $ILR_s$ at $R <$ 1.7$R_{CR}$ (the 25/55 model) implies that stars
will be pushed into the relevant annulus from lower radii than when
$ILR_s$ concurs with the physical location of $OLR_b$. In the case
$ILR_s$ occurs at $1.7 R_{CR}<R <2.3 R_{CR}$, (the 18/55 model), a
part of the annulus will be depleted at the cost of the parts at radii
around 2.3$R_{CR}$. Thus, for the 25/55 model, more stars will be
entering our annulus from lower energies than in the other two
models. Now, in a smooth unperturbed background potential, stars at
lower radii are also more energetic than those at higher radii. This
implies that in the absence of resonances due to imposed
perturbations, there would have been more high energy stars recorded
for the 25/55 case than in the 18/55 or 21/55 models.

This situation is of course challenged once the perturbations are
introduced - in particular, proximity to resonance overlap indicates
enhanced chaoticity in the recorded orbits. The relative excess in the
energy of the recorded orbits, as implied by the 25/55 model is
surpassed, more at higher energies than lower, by the strength of
chaos that is a signature of the resonance overlap case. This explains
the relative trends in chaos strength that is noticed in the different
models (Figure~4).

We conclude that the observed phase space structure in the Solar
neighbourhood (particularly the splitting of the Hyades-Pleiades mode)
is to a large extent, chaos induced. But this chaos does not
necessarily have to be triggered by resonance overlap (in
contradiction to what Quillen, 2003 suggested). In fact, the presence
of {\it chaos is found to be actuated by the spiral potential}.
We say this since our results indicate that the {\it bar potential
alone is insufficient in producing chaos}. This is in contradiction to
the suggestion by Fux (2001). The bar that was used in the modelling
in Paper~I (our models) imposes a field of 3.6$\%$ of that
of the background disk, nearly half of what was used by Fux
(2001). Thus, it may be argued that it is this low a bar strength that
was incapable of heating the disk enough; after all, as shown in
Chakrabarty (2004), disk heating increases rapidly with increases in
bar strength.

\section{Conclusions}
\noindent
In this work, we have presented a neat way of quantifying chaos that
shows up in models of the local phase space. This work needs to be
buttressed in the future with more sophisticated models that span all
six phase space dimensions and account for the Galactic halo as
well. This objective estimation and classification of orbits into
strongly chaotic, weakly chaotic and regular, allows us to understand
the local phase space in greater details than has been possible
before. We implement this technique on models of the Solar
neighbourhood to conclude that all models that include the spiral
pattern exhibit chaoticity and this nature of the local phase space is
advanced as an important contributor to the formation of the observed
phase space structure. We advance this technique as a blueprint for
evaluating the degree of chaos present in kinematic samples that would
be collated in the near future by GAIA.

\begin{acknowledgements}
\noindent
DC is supported by a Royal Society Dorothy Hodgkin Fellowship. IVS is
supported by the Tomalla Foundation.
\end{acknowledgements}

\end{document}